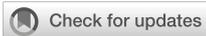

# Oxygenic photosynthetic responses of cyanobacteria exposed under an M-dwarf starlight simulator: Implications for exoplanet's habitability

Mariano Battistuzzi[1,2]*, Lorenzo Cocola[3], Riccardo Claudi[4], Anna Caterina Pozzer[1,4], Anna Segalla[1], Diana Simionato[1†], Tomas Morosinotto[1,2], Luca Poletto[3] and Nicoletta La Rocca[1,2]*

[1]Department of Biology, University of Padua, Padua, Italy, [2]Center for Space Studies and Activities (CISAS), University of Padua, Padua, Italy, [3]National Council of Research of Italy, Institute for Photonics and Nanotechnologies (CNR-IFN), Padua, Italy, [4]National Institute for Astrophysics, Astronomical Observatory of Padua (INAF-OAPD), Padua, Italy

**Introduction:** The search for life on distant exoplanets is expected to rely on atmospheric biosignatures detection, such as oxygen of biological origin. However, it is not demonstrated how much oxygenic photosynthesis, which on Earth depends on visible light, could work under spectral conditions simulating exoplanets orbiting the Habitable Zone of M-dwarf stars, which have low light emission in the visible and high light emission in the far-red/near-infrared. By utilizing cyanobacteria, the first organisms to evolve oxygenic photosynthesis on our planet, and a starlight simulator capable of accurately reproducing the emission spectrum of an M-dwarf in the range 350–900 nm, we could answer this question.

**Methods:** We performed experiments with the cyanobacterium *Chlorogloeopsis fritschii* PCC6912, capable of Far-Red Light Photoacclimation (FaRLiP), which allows the strain to harvest far-red in addition to visible light for photosynthesis, and *Synechocystis* sp. PCC6803, a species unable to perform this photoacclimation, comparing their responses when exposed to three simulated light spectra: M-dwarf, solar and far-red. We analysed growth and photosynthetic acclimation features in terms of pigment composition and photosystems organization. Finally, we determined the oxygen production of the strains directly exposed to the different spectra.

**Results:** Both cyanobacteria were shown to grow and photosynthesize similarly under M-dwarf and solar light conditions: *Synechocystis* sp. by utilizing the few photons in the visible, *C. fritschii* by harvesting both visible and far-red light, activating the FaRLiP response.

**Discussion:** Our results experimentally show that an M-dwarf light spectrum could support a biological oxygen production similar to that in solar light at the tested light intensities, suggesting the possibility to discover such atmospheric biosignatures on those exoplanets if other boundary conditions are met.

# Introduction

In 25 years of surveys, more than 5000 confirmed exoplanets were found (https://exoplanets.nasa.gov/). Among them, small, rocky terrestrial-like exoplanets were discovered (Tinetti et al., 2013), mainly laying in the Habitable Zone (HZ, Kasting et al., 1993) of their host stars. Terrestrial-like exoplanets are rather common around M-dwarfs (M class Main Sequence stars, or also Red Dwarf stars) (Bonfils et al., 2013; Kopparapu et al., 2013; Hsu et al., 2020), very faint and cool stars (Pecaut and Mamajek, 2013) with most of their emission centred in the far-red (700 – 750 nm) and infrared (750 – 1000 nm), rather than in the visible (400 – 700 nm) like the Sun (G class Main Sequence star, or G2V). Despite high stellar activity during their early stages of life and stellar flares (Kiang et al., 2007; Segura et al., 2010; France et al., 2013; Cuntz and Guinan, 2016), orbiting exoplanets could remain habitable (O'Malley-James and Kaltenegger, 2017; Schwieterman et al., 2018). Moreover, because of the longevity of M-dwarfs (Adams and Laughlin, 1997), exoplanets in their HZ seem the best candidates to search for life beyond Earth. The closest exoplanet known to be orbiting an M-dwarf in its HZ, Proxima Centauri b (Anglada-Escudé et al., 2016) is 4.2 light-years away, meaning that the search for possible lifeforms will have to be made through detection of *biosignatures*: substances, molecules or patterns of biological origin (Seager et al., 2012; Schwieterman et al., 2018; Claudi and Alei, 2019). On Earth, the most known and referenced biosignature is oxygen ($O_2$) and its photoproduct ozone ($O_3$), derived from the oxygenic photosynthesis (OP) performed by cyanobacteria, algae and plants. This process is unique with respect to other photosynthetic processes (e.g., anoxygenic photosynthesis) due to the utilization of $H_2O$ as a donor of electrons, an extremely abundant, yet very stable and very difficult molecule to oxidize. Thanks to OP, $O_2$ which was originally in traces reached 21% by volume of the Earth's atmosphere, becoming a dominant gas. Abiotic processes on Earth account for less than 1 ppm of atmospheric $O_2$ and are considered negligible Harman et al. (2015). For exoplanets, $O_2$ could be considered a robust biosignature, even if care must be taken for possible false positives arising from photochemical reactions that have abiotic $O_2$ as a by-product (e.g., $H_2O$ and $CO_2$ photoionization). A detailed discussion of several possible false positives is presented in Harman et al., 2015. On the other side, life on exoplanets could also not use oxygenic photosynthesis, and rely for example on anoxygenic photosynthesis which utilizes $H_2$, $H_2S$, $Fe^{2+}$ as electron donors instead of $H_2O$. These molecules however are not as widely distributed as $H_2O$ is, therefore anoxygenic photosynthesis would be less productive, leading to a harder detectability of possible by-product gasses (Kiang et al., 2007; Schwieterman et al., 2018). To evaluate the possibility, for an exoplanet orbiting the HZ of an M-dwarf, to undergo an oxygenation event as happened on Earth and reach through it detectable levels of $O_2$ in its atmosphere, one way is to select and study, through simulations, suitable organisms among terrestrial ones, keeping in mind some of the environmental characteristics expected to be present on those planets. Organisms need to be grown phototrophically under low light and/or far-red light conditions, and under continuous light irradiation, as exoplanets orbiting M-dwarfs can be tidally locked (Barnes, 2017). It's also preferable to select among Earth's extremophiles since very little information is available on the surface environmental conditions of those planets. Cyanobacteria fulfil all these requisites. These organisms drove important evolutionary events on Earth: i) being responsible for the emergence of oxygenic photosynthesis, which led to the accumulation of $O_2$ in the Earth's atmosphere that caused the Great Oxidation Event and later promoted the evolution of more complex lifeforms (Lyons et al., 2014); ii) evolving the capability to withstand extreme environmental conditions, which allowed them to spread everywhere on Earth and become major primary producers from an ecological point of view (Gan and Bryant, 2015); iii) having high plasticity of their photosynthetic apparatus, which made them able to live in habitats characterized by deeply different light intensities and spectra (Gutu and Kehoe, 2012); iv) acquiring the ability to fix $N_2$ beside $CO_2$, thus causing significant changes both in the Earth's atmosphere and in the $N_2$ and $CO_2$ geochemical cycles (Schirrmeister et al., 2016). Oxygenic photosynthesis employs two types of pigment-protein complexes, photosystems I and II (PSI, PSII), both composed of a reaction centre, which converts light into chemical energy, and an antenna, which harvests and funnels the light energy into the reaction centre (Bryant and Canniffe, 2018). The reaction centres are highly conserved in all oxygenic photosynthetic organisms and link chlorophyll *a* and carotenoids, which have major absorption bands in the blue (400-500 nm) and red (600-700 nm) wavebands. Antenna systems, instead, are diversified in the different taxa depending on the specific light-harvesting pigments associated with proteins, conferring them a distinctive capability to absorb photons in the visible portion of the solar spectrum, generally defined as Photosynthetically Active Radiation or PAR (400 – 700 nm). Most oxygenic photosynthetic organisms absorb blue and red light, while cyanobacteria are able to harvest a larger spectrum of wavelengths thanks to antenna complexes called phycobilisomes (PBS), made by different species-specific combinations of phycobiliproteins (PBPs) such as phycoerythrin (PE), phycocyanin (PC) and allophycocyanin (AP), which have peak absorptions at roughly 565, 615 and 655 nm (Bryant and Canniffe, 2018). Moreover, even if most cyanobacteria use visible light-absorbing pigments to perform oxygenic photosynthesis, it has been shown that some species are able to harvest photons in the far-red region (700 – 800 nm), thanks to the synthesis of alternative types of chlorophylls beside chlorophyll *a*. They live indeed in habitats depleted of visible light and enriched in far-red, due to physical light scattering and attenuation (sands, caves, water columns) or by absorption of other photosynthetic organisms living on top of them (plant canopies, microbial mats, stromatolites) (Gan and Bryant, 2015). They evolved to utilize this light through the so-called Far-Red Light Photoacclimation (FaRLiP). This process was first discovered in *Leptolyngbya* sp. JSC-1 and then identified in a dozen species of terrestrial cyanobacteria (Gan et al., 2014a; Gan et al., 2014b; Li et al., 2016) and seems to be widely distributed in natural environments (Zhang et al., 2019; Antonaru et al., 2020; Kühl et al., 2020). These cyanobacteria deeply modify their photosynthetic machinery to absorb far-red light: they synthesize chlorophylls *d* and *f* (Chl *d*, Chl *f*) (Chen, 2019), but also far-red-absorbing forms of AP. They substitute core subunits of the PSI, PSII and PBS with far-red paralogs and modify the morphology of the PBS antennae to better absorb light above 700 nm, as can be observed in their *in vivo* absorption spectra (Gan et al., 2014a; Gan and Bryant, 2015; Li et al., 2016). With respect to other oxygenic photosynthetic organisms, cyanobacteria are thus characterized by high photosynthetic plasticity and can integrate the absorption of chlorophylls and carotenoids expanding the wavelengths of light they can harvest for photosynthesis. Recently (Claudi et al., 2021) we obtained preliminary important data exposing few cyanobacterial species to a simulated M-dwarf light spectrum, demonstrating their survival and growth under this exotic light condition. Based on chlorophyll fluorescence measurements, we also showed their capability to perform photosynthesis. We here exposed two species of cyanobacteria, one, *Chlorogloeopsis fritschii* PCC6912, able to perform FaRLiP and another, *Synechocystis* sp. PCC6803, unable to perform it, to different light spectra. Their growth, photosynthetic apparatus organization and changes in pigment composition, as well as their oxygenic photosynthetic evolution rate, were investigated under the simulated M-dwarf light spectrum along with two control conditions: a simulated solar light spectrum, to assess the behaviour of the cyanobacteria in Earth-like conditions, and a far-red light spectrum, to check the activation of the FaRLiP response. Results demonstrate that both cyanobacteria grown under a simulated M-dwarf spectrum acclimate in different ways, growing and evolving oxygen as efficiently as in a solar simulated spectrum, despite the lesser amount of visible light. They could therefore produce atmospheric biosignatures detectable from remote.

## Materials and methods

### Experimental design

The research has been carried out using two different cyanobacterial strains, respectively able and unable of FaRLiP response. For each strain, short-term and long-term (3 and 21 days respectively) acclimation experiments were performed, exposing 4 independent culture replicates (N = 4) in flasks to three different light sources (one for starlight, M7, one for far-red light, FR, one for solar light, SOL), under terrestrial atmospheric composition. A total of 24 flasks per strain was utilized. The different timings were selected based on preliminary experiments with the FaRLiP strain, which had shown the minimum time required to evidence the beginning of an acclimation response to FR (3 days) and the minimum time required to have a full acclimation response to FR (21 days), based on *in vivo* absorption measurements (Figure S1). To better characterize the initial phases of the FaRLiP process, *in vivo* absorption spectroscopy, chl *d* and *f* detection through HPLC and 77K fluorescence spectroscopy analyses were carried out at 3 days for each culture replicate. 21 days experiments have been performed to study the responses of each strain to long-term exposure under the selected light sources. The growth curves have been obtained by sampling the cultures in sterility and dim light, by measuring the optical density at 750 nm 3 days a week. Moreover, at the beginning and the end of the 21 days experiments, dry weight, pigments content, *in vivo* absorption spectroscopy, chl *d* and *f* detection through HPLC, 77K fluorescence spectroscopy, and optical microscopy analysis were performed to characterize the acclimated cultures.

A further experiment has been carried out to determine the $O_2$ production capabilities of the strains once long-term acclimated to the three light conditions. Similarly to the previous experiment, 3 independent culture replicates (N = 3) have been grown in flasks for 21 days, exposed to the three different light sources. At the end of the acclimation time, for each culture replicate, the optical density, the chlorophyll *a* content, and the *in vivo* absorption spectrum were checked. For each organism and light condition tested, equal parts of the cultures were pooled together to measure the $O_2$ production of the acclimated strains directly under the three different light sources, by utilizing growth chambers (Atmosphere Simulator Chamber, ASC) equipped with $O_2$ sensors.

### Experimental setup

The experimental setup is composed of 3 different temperature-controlled cabinets (Figure 1A), each hosting a different light source which illuminates a flatbed, that can be used to cultivate organisms in flasks under a terrestrial atmosphere. Upon necessity, the flatbed can host the growth chamber (ASC, Figures 1B, C), which can be utilized to expose the cultures simultaneously to different atmospheric compositions and irradiations from the different light sources, allowing also to monitor the $O_2$ released in the ASC in real-time. For the experiments, three different light sources were used (Figure 2). Each light spectrum was set at 30 µmol $m^{-2}$ $s^{-1}$ in the range between 380 and 780 nm as checked through a spectrometer (LI-COR 180, LI-COR). The M-dwarf light spectrum (M7) was simulated using the Star Light Simulator (SLS), extensively described in (Battistuzzi et al., 2020; Claudi et al., 2021). The Solar spectrum (SOL) was reproduced through a custom-made simulator which presents a circular 50 mm diameter plate on which are soldered the LEDs, arranged into 4 channels (Table S1). A mixing chamber allows the homogenization of the light spectrum emitted by the simulator. The Far-Red spectrum

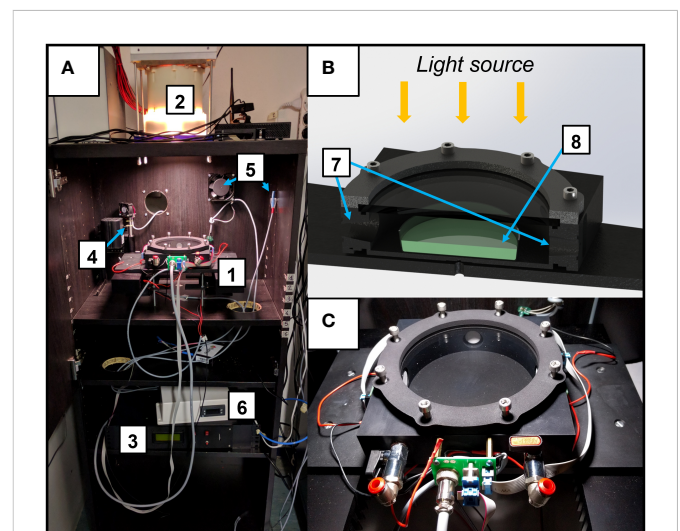

FIGURE 1
The experimental setup utilized for the experiments. **(A)** Temperature-controlled cabinet. Here is shown the cabinet with the Star Light Simulator (SLS) mounted on top and the Atmosphere Simulator Chamber (ASC) positioned in the flatbed under it; the major elements of the setup are highlighted with numbers: 1. ASC; 2. Light source (here is the SLS); 3. Control software for the ASC; 4,5. Temperature control elements of the cabinet (heater, fan, temperature probe); 6. Main temperature control of the cabinet. **(B)** Cutaway diagram of the ASC; 7. $CO_2$ and $O_2$ sensors housings; 8. Petri dish with cyanobacteria on it; **(C)** close-up of the ASC.

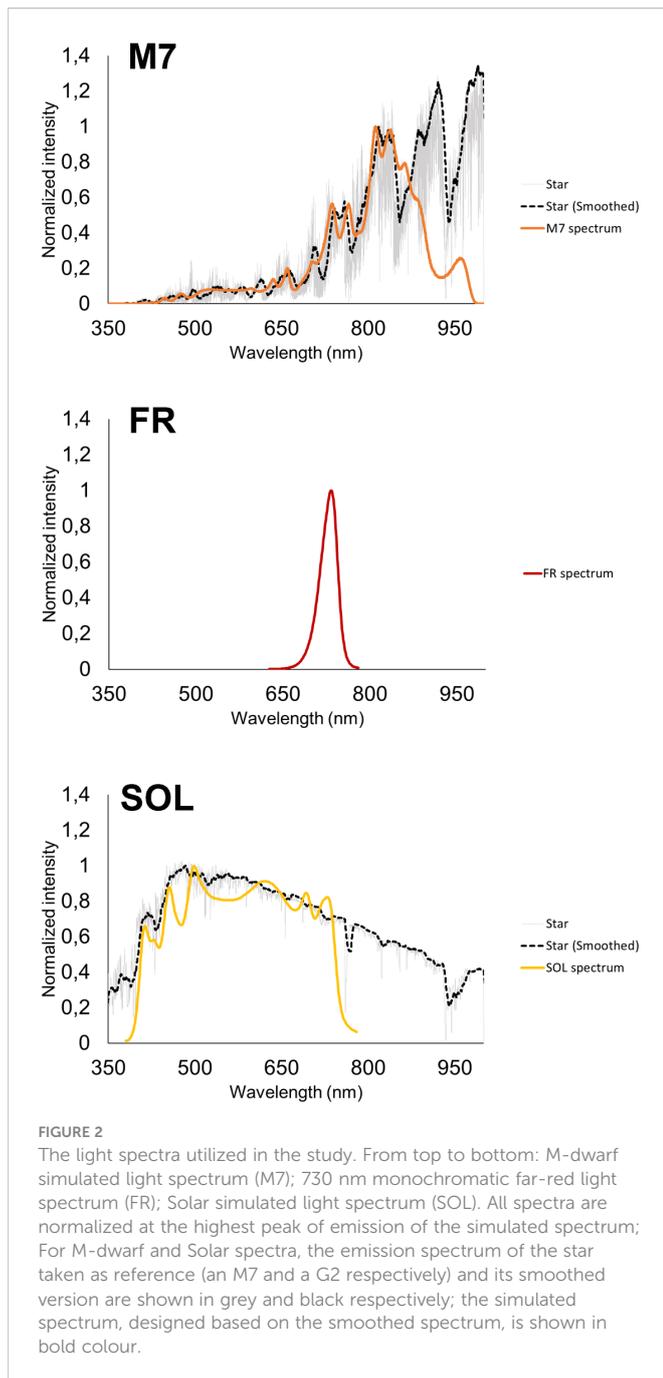

FIGURE 2

The light spectra utilized in the study. From top to bottom: M-dwarf simulated light spectrum (M7); 730 nm monochromatic far-red light spectrum (FR); Solar simulated light spectrum (SOL). All spectra are normalized at the highest peak of emission of the simulated spectrum; For M-dwarf and Solar spectra, the emission spectrum of the star taken as reference (an M7 and a G2 respectively) and its smoothed version are shown in grey and black respectively; the simulated spectrum, designed based on the smoothed spectrum, is shown in bold colour.

(FR) was reproduced through a custom-made illuminator made of a rectangular 70 mm x 70 mm plate on which are soldered 24 LEDs emitting at 730 nm (XP-E2 FAR RED, CREE) distributed in 5 parallel lines. To monitor the $CO_2$ and $O_2$ exchanges of the acclimated strains, three ASC were utilized (Figures 1B, C). These chambers are based on a previous version of the ASC (Battistuzzi et al., 2020), are made of black anodized aluminium, have a volume of 0.45 L, a diameter of 125 mm and a height of 33 mm. The chambers allow us to carry on experiments on one sample at a time in an open Petri dish with up to 70 ml of liquid culture. On top of the chambers, a lid with a Borofloat window allows the illumination of the samples. On the side of the chambers, three 24 mm and 5 mm holes take place. Two 24 mm holes are occupied by $CO_2$ sensors (CO2M-20 and CO2M-100, SST sensing), the other one of the same size by an $O_2$ sensor (LOX-O2-

S, SST sensing) (Table S2). The two 5 mm holes can be connected to pipe fittings and different flow meters and needle valves, to flux a desired atmosphere inside the chambers. Data collected from the sensors are stored in a microSD card for later use. The working pressure and temperature ranges of the ASC go respectively from 0 to ~140 kPa and from 15 to 40 ± 0.5°C.

## Cultivation conditions

The selected strains, *Chlorogloeopsis fritschii* PCC6912 (hereafter, PCC6912) and *Synechocystis* sp. PCC6803 (hereafter, PCC6803), acquired from the Pasteur Culture Collection (PCC, France), were maintained in liquid cultures in a climatic chamber at 30 ± 0.5°C, exposed to a continuous cool white fluorescent light of 30 µmols m$^{-2}$ s$^{-1}$ (L36W-840, OSRAM) under a terrestrial atmospheric composition. Both strains were maintained in a BG-11 liquid medium (Rippka et al., 1979).

Depending on the experiment, cells from maintenance cultures were pre-inoculated in up to 4 independent flasks at an optical density at 750 nm (OD$_{750}$) of 0.2 in a final volume of 100 mL and were exposed to the SOL light spectrum at an intensity of 30 µmols m$^{-2}$ s$^{-1}$ (380 – 780 nm) until they reached exponential phase: OD$_{750}$ of about 0.9 for 6912 and OD$_{750}$ of about 0.5 for 6803. Each flask of the pre-inoculum was then gently centrifuged at 3500 g for 5 min and resuspended to an OD$_{750}$ of 0.2 with fresh BG-11 medium, split into 3 new flasks in a final volume of 50 mL, each exposed to one of three different spectra (an M-dwarf light spectrum, a far-red spectrum, a solar spectrum) at a total light intensity of 30 µmols m$^{-2}$ s$^{-1}$ (380 – 780 nm), inside the temperature-controlled cabinet set at 30 ± 0.5°C. All inocula were performed under sterile conditions. The sterility of the cultures at the end of the experiments was confirmed through microscopy, and cell vitality was assessed through optical and fluorescence microscopy (Figures 3, S2).

## Dry weight determination

To determine the dried biomass concentration, 10 mL of culture were diluted 1:3 with de-ionized water and filtered with a vacuum flask on 0.45 µm nitrocellulose filters (Sigma-Aldrich), previously dried in a heater at 70°C for at least 3 h and weighted. Filters with cyanobacteria were put again in the heater at 70°C to dry and weighed at least after 24 h. Dry weight (DW) was calculated as follows:

$$Dry\ Weight\ \left[\frac{g}{L}\right] = \frac{(Filter\ with\ cyanobacteria\ [g] - Empty\ filter[g])}{Volume\ of\ culture\ filtered\ [mL]} \times 1000$$

## *In vivo* absorption spectra measurements

For *in vivo* absorption analysis, 1.5 mL of culture were centrifuged twice at 1400 g for 5 min (Centrifuge Sigma 3K15). The supernatant was discarded, and the pellet was homogenized with a pestle and then resuspended in 600 µL of fresh BG-11 medium. The obtained suspension was used for the spectrophotometric analyses (Spectrophotometer, Agilent Cary 300 UV-VIS), using quartz cuvettes and exposing their opaque side to the ray to correct scattering (Gan et al., 2014a).

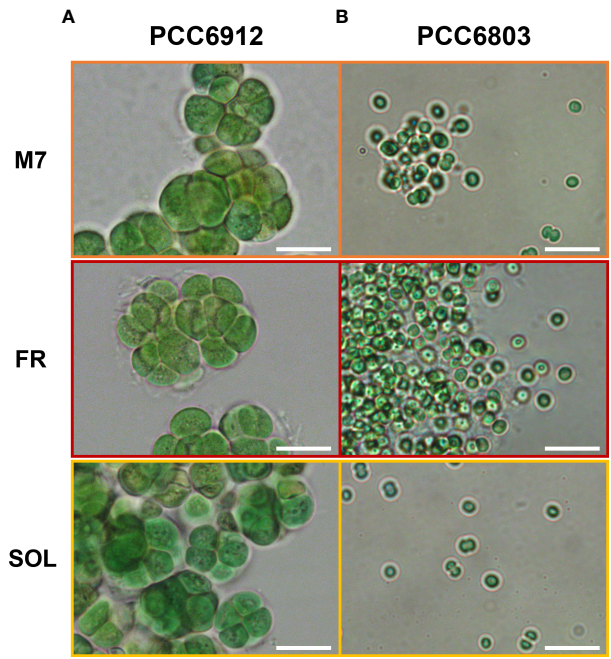

FIGURE 3
Bright-field optical microscopy of PCC6912 **(A)** and PCC6803 **(B)** samples after 21 days of exposure to the selected light spectra at higher magnification. Images were obtained through an optical microscope (Leica DM6B, Leica, Wetzlar, Germany), utilizing white lights, and maintaining the acquisition settings for each sample. Orange: M7; red: FR; yellow: SOL. Scale bars: 10 µm.

## Chlorophyll *a* and carotenoids content determination

To evaluate the lipophilic pigment content of the strains, 2 mL of culture were centrifuged for 10 min at 17500 g. The supernatant was discarded, and the pellet was solubilized in 1 mL of DMF (N,N′-dimethylformamide). Samples were kept at least for 24 h at 4°C in the dark, to allow the extraction of lipophilic pigments. Pigment spectra were recorded with a spectrophotometer (Cary 300 UV-Vis, Agilent) by using glass cuvettes. The concentrations of chlorophyll *a* and total carotenoids were obtained using the Moran equations for DMF (Moran, 1982).

## Phycobiliproteins content determination

To evaluate the hydrophilic pigment content of the samples, 4 mL of cultures were centrifuged at 17500 g for 5 min at 4°C. The supernatant was discarded and to the pellet were added 20 µL of extraction buffer (EB, $Na_2HPO_4$ 0,01M, NaCl 0,15M) and an equal amount of glass beads (150-212 µm, SIGMA). Cell pellets were broken up with three cycles of Bead-Beater (Biospec Products), composed of 10 s at 3500 OPM (oscillations per minute) followed by 30 s in ice. To the lysate were then added 180 µL of EB and another cycle of Bead-Beater was repeated for 4 s. The sample was then centrifuged two times at 20.000 g, for 5 min, at 4°C, and the supernatant was moved to a new microvial. A second extraction was performed by adding to the pellet 200 µL of EB; the pellet was solubilized through vortexing and then centrifuged as before. Further extractions were repeated until a transparent supernatant and a light green pellet were observed. All supernatants were pulled and the final phycobiliprotein extract was kept at -20°C in the dark until further analyses. Spectrophotometric analysis was performed as described for the lipophilic pigments. The concentrations of phycocyanin and allophycocyanin were obtained using Bennet and Bogorad equations (Bennett and Bogorad, 1973).

## HPLC analysis

For liquid chromatographic analyses, 4 mL of culture were centrifuged twice at 17500 g, for 5 min, at 4°C. The supernatant was discarded and glass beads (150-212 µm, SIGMA) were added to the pellet, together with 20 µL of acetone 90%. Samples were disrupted with three cycles of rupture for 1 s at 3500 OPM (oscillations per minute) through a Bead-Beater (Biospec Products) followed by 30 s in ice. To the mixture were then added 180 µL of acetone 90% and the rupture was repeated once for 4 s. The samples were then centrifuged twice at 20.000 g, for 5 min, at 4°C, and the supernatants were kept aside. To ensure complete extraction of the pigments, a second extraction was performed by adding to the pellets 800 µL of acetone 90%, and by vortexing and subsequently centrifuging the mixture as before. Extracts obtained this way were added to the previous extraction to obtain about 1 mL of final pigment extract and kept at -20°C until analyses. For the analyses, an Agilent 1100 series LC system was utilized. The stationary phase was a column (length 250 mm, diameter 4 mm) internally filled with 5 μm silica particles coated by C-18 atom chains (Merck Lichrospher 100 RP). The mobile phase was represented by two solutions: A) methanol: acetonitrile: H₂O milliQ (42:33:25 ratio); B) methanol: acetonitrile: etilacetate (50:20:30 ratio). Solutions A and B were eluted in the column following the protocol reported in Gan et al., 2014a, optimized for the detection of Chl *d* and *f*. During each run, the detector registered the absorption levels of each pigment eluted from the column when illuminated by a 705 nm source light.

## 77 K temperature fluorescence analyses

To analyze the fluorescence emission spectra of the tested samples, 1 mL of culture was centrifuged at 1500 g for 5 min and the obtained pellet was resuspended in glycerol 60% (w/v) and Hepes (pH 7.5, 10mM). The mixture was rapidly frozen in liquid nitrogen and kept at -80°C up until analysis. Analyses were performed using a fluorimeter (Cary Eclipse Fluorescence Spectrophotometer, Agilent) and exciting samples at 440 nm, to evaluate variation in the organization of the photosynthetic apparatus related to chlorophylls. Emission spectra were evaluated in the range between 600 and 800 nm (Gan et al., 2014a).

## Gas exchange measurements

The cyanobacteria photosynthetic rate was measured for 24 h when the organisms, acclimated for 21 days to the different spectra, were directly exposed to the respective growth light conditions inside the ASC. Before these measurements, cell cultures were grown in flasks in the cabinets as for the other experiments, then they were used after checking that their physiological responses were those expected. For each strain and light condition, samples were pooled together, due to the impossibility to evaluate more than one sample at a time in the ASC. Then, 70 mL from the pool were directly poured into previously sterilized glass Petri dishes and immediately closed inside the ASC for 24 h under the different light sources at 30 μmols m$^{-2}$ s$^{-1}$ into an initial atmospheric composition of air + 5% $CO_2$ at 1 atm, 30°C. A concentration of 5% $CO_2$ is used to sustain the photosynthetic activity of the strains throughout the experiments; otherwise, cyanobacteria in the closed chambers will suddenly deplete terrestrial $CO_2$ concentrations (Battistuzzi et al., 2020). When the ASC were positioned inside the temperature-controlled cabinet, the latter was kept at a slightly higher temperature than that of the ASC itself. This helps to avoid condensation phenomena on the top of the ASC glass windows and improves the accuracy of the ASC temperature control loops. Raw data of $O_2$ levels registered from the $O_2$ sensors were elaborated through Matlab (MathWorks) to obtain the $O_2$ production expressed in micromoles by applying the ideal gas law (Battistuzzi et al., 2020).

## Statistical analyses

Statistical analyses were performed through the software Graph Pad Prism v7.0 (GraphPad software). Data were calculated as the mean ± standard deviation of 4 different biological replicates and the comparison between different light conditions (M7, SOL, FR) for the same strain was carried on with the one-way ANOVA technique (assuming the Gaussian distribution of data) followed by Tukey's multiple comparison test (significance was set at p< 0.05).

## Results

### The growth of FaRLiP and non-FaRLiP cyanobacteria is not limited under M7 with respect to SOL

The growth of both cyanobacteria strains exposed to the three light conditions was followed for 21 days, measuring the OD 3 times a week (Figure 4). The growth curves of cells under M7 and SOL were almost superimposable within each strain, demonstrating the capability of FaRLiP and non-FaRLiP cyanobacteria to grow similarly under these two light spectra. In FR light, only PCC6912 was able to grow as expected. This was confirmed by the maximum growth rates data ($\mu_{max}$, OD days$^{-1}$) calculated from the third to the seventh day for PCC6912 and from the second to the fourth day for PCC6803 for each light condition. PCC6912 exposed to SOL and M7 showed a similar OD increase over time ($\mu_{max}$ of 0.21 ± 0.03 days$^{-1}$ and 0.21 ± 0.05 days$^{-1}$, respectively), while in FR its growth was clearly slower ($\mu_{max}$ of 0.07 ± 0.05 days$^{-1}$). PCC6803 growth was also vigorous both in SOL and M7 ($\mu_{max}$ of 0.48 ± 0.05 days$^{-1}$ and 0.36 ± 0.05 days$^{-1}$, respectively), instead, just the survival of cells was recorded in FR ($\mu_{max}$ of 0.00 ± 0.05 days$^{-1}$) (Figures 3, S2). The two organisms reached different OD values at the end of the experiments due to their different replication time and cell dimension, which influence OD measurements. Overall, PCC6803 demonstrated a faster growth in OD with respect to PCC6912, even if final biomass quantifications were similar for both strains acclimated to M7 and SOL.

These responses were also clearly visible by looking at the pigmentation of the cultures of both strains at 21 days (Figure 5). Biomass quantifications were mostly consistent with OD (Figure 4C). PCC6912 cells registered a significant biomass increase under SOL with respect to M7, and an increase in biomass also for FR. PCC6803 cells instead registered similar increases of biomass under SOL and M7 and no significant growth in FR. Overall, PCC6803 demonstrated a faster growth in OD with respect to PCC6912, even if final biomass quantifications were similar for both strains acclimated to M7 and SOL. After cultivation, the pigment concentration of chlorophyll *a* (Chl *a*), total carotenoids (Car), phycocyanin (PC) and allophycocyanin (AP) of the strains was assessed (Table 1). All pigments quantified over dry weight for PCC6912 were significantly higher in M7 than in SOL and FR. In PCC6803 instead Chl *a* content was significantly higher in M7 than in SOL and FR, while Car content was significantly higher in FR than in M7 and SOL. Regarding phycobiliproteins, PC and AP contents were significantly higher in M7 than in SOL and FR.

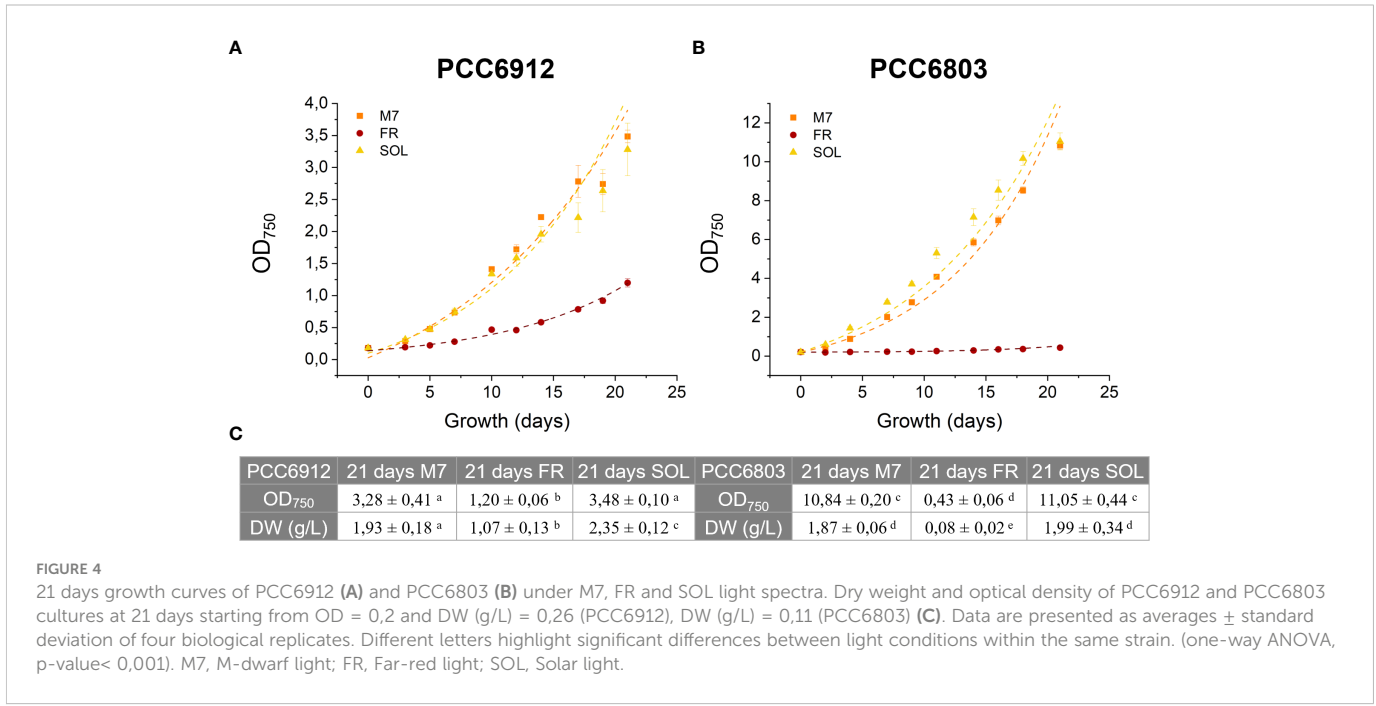

FIGURE 4

21 days growth curves of PCC6912 **(A)** and PCC6803 **(B)** under M7, FR and SOL light spectra. Dry weight and optical density of PCC6912 and PCC6803 cultures at 21 days starting from OD = 0,2 and DW (g/L) = 0,26 (PCC6912), DW (g/L) = 0,11 (PCC6803) **(C)**. Data are presented as averages ± standard deviation of four biological replicates. Different letters highlight significant differences between light conditions within the same strain. (one-way ANOVA, p-value< 0,001). M7, M-dwarf light; FR, Far-red light; SOL, Solar light.

## The FaRLiP response in *C. fritschii* is a slow process under M7 with respect to FR

The activation of FaRLiP was first assessed through *in vivo* absorption spectra measurements. Spectra were recorded after 3 days for every condition tested (Figure S3). PCC6912 showed a very small *in vivo* absorption of light beyond 700 nm under FR after 3 days. Under M7 and SOL, however, no signs of such absorption were observable. PCC6803 showed no additional *in vivo* pigment absorption beyond 700 nm for any of the tested conditions after 3 days, confirming the inability of the strain to utilize wavelengths different from those in the visible range. After 21 days though, the *in vivo* absorption of the strains changed, depending on the species

(Figure 6). PCC6912 *in vivo* absorption of far-red light became evident both for FR and M7 long-term acclimated cells, showing absorption of light beyond 700 nm, while in SOL ones this was never detected. Contextually, an increase in the PC/Chl *a* ratio in SOL and M7 with respect to FR was observed. PCC6803 showed no additional peaks of *in vivo* absorption beyond 700 nm as expected, but it was instead observed an increase in the PC/Chl *a* ratio in FR with respect to the M7 and SOL, confirming the pigment concentration analyses (Table 1). To further investigate the FaRLiP response in the cultures exposed to FR and M7, HPLC analyses were conducted at 3 and 21 days (only 21 days measurements are shown here, for 3 days measurements see Figure S4) (Figure 7). PCC6912 pigment extracts showed that chlorophylls *d* and *f* were detectable both in FR and M7 after 3 and 21 days, while for PCC6803 pigment extracts chlorophylls *d* and *f* were never detected. To determine changes in the organization of the photosynthetic apparatus following the acclimation response of the strains exposed to M7, 77 K fluorescence measurements were made after 3 and 21 days (Figures 8, 9). PCC6912 exhibited different emission spectra at 440 nm for each light condition tested. At 3 days, SOL acclimated samples showed a single emission peak at 725 nm (white light PSI, WL-PSI) and minor peaks at 685 and 695 nm (white light PSII, WL-PSII). FR acclimated samples instead exhibited a reduction in the emission peak at 725 nm and an additional peak at 740 nm (far-red light PSI, FRL-PSI). Changes were also observed in the minor emission peaks, with the reduction of peaks at 685 and 695 nm. Notably, M7 acclimated samples also exhibited a very small shift in the PSI emission fluorescence. PSII emission peaks were instead similar to those in SOL acclimated samples. On the contrary, no variation in the emission peak of PSI occurred in the photosynthetic apparatus of PCC6803 under any of the light conditions tested after 3 days. The organization of the photosynthetic apparatus of both cyanobacteria changed dramatically upon acclimation to both M7 and FR after 21 days (Figure 9). After 21 days, PCC6912 samples acclimated to FR presented an emission peak of PSI which was

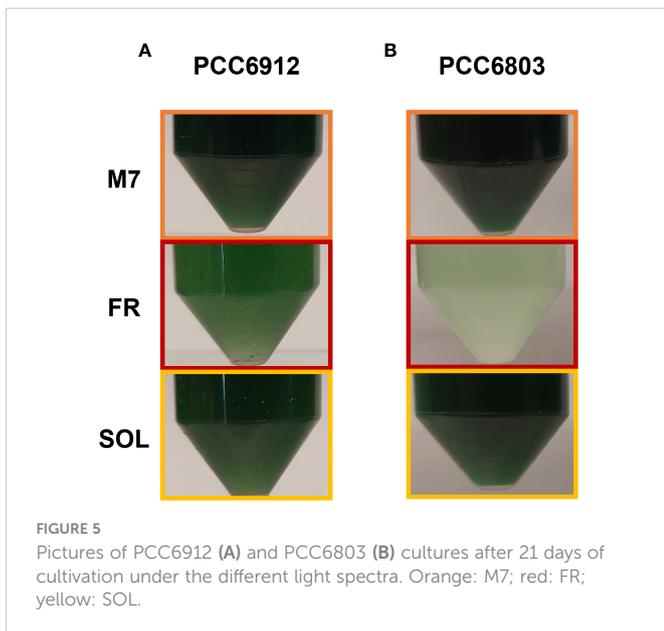

FIGURE 5
Pictures of PCC6912 **(A)** and PCC6803 **(B)** cultures after 21 days of cultivation under the different light spectra. Orange: M7; red: FR; yellow: SOL.

TABLE 1 Pigment content of PCC6912 and PCC6803 cultures at day 21.

| PCC6912 | 21 days M7 | 21 days FR | 21 days SOL |
|---|---|---|---|
| Chl a (mg/gr DW) | 12,40 ± 1,37[a] | 9,82 ± 0,58[b] | 9,35 ± 0,82[b] |
| Car (mg/gr DW) | 2,74 ± 0,19[a] | 2,10 ± 0,13[b] | 1,76 ± 0,16[c] |
| PC (mg/gr DW) | 73,79 ± 4,14[a] | 37,54 ± 4,44[b] | 44,73 ± 8,27[b] |
| AP (mg/gr DW) | 40,74 ± 2,54[a] | 19,57 ± 2,94[b] | 22,58 ± 5,00[b] |
| PCC6803 | 21 days M7 | 21 days FR | 21 days SOL |
| Chl a (mg/gr DW) | 8,64 ± 0,20[a] | 4,33 ± 0,47[b] | 6,31 ± 0,75[c] |
| Car (mg/gr DW) | 2,22 ± 0,07[a] | 2,75 ± 0,37[b] | 2,24 ± 0,24[b] |
| PC (mg/gr DW) | 104,36 ± 7,91[a] | 62,8 ± 7,69[b] | 57,97 ± 12,02[b] |
| AP (mg/gr DW) | 32,97 ± 6,10[a] | 13,00 ± 3,05[b] | 13,45 ± 4,95[b] |

Chl a, Chlorophyll a; Car, total carotenoids; PC, phycocyanin; AP, allophycocyanin; M7, M-dwarf light; FR, Far-red light; SOL, Solar light. Data are presented as averages ± standard deviation of four biological replicates. Different letters highlight significant differences for the same parameter measured within a single strain under different light conditions. (one-way ANOVA, p-value< 0,001).

completely shifted towards 740 nm; no clear emission peaks were detected from PSII. M7 acclimated samples presented instead a reduction in the emission peak at 725 nm and the shift of the emission towards 740 nm, presenting a spectrum almost identical to that of 3 days FR acclimated samples, except for PSII emission, which was slightly higher in these samples. SOL acclimated samples presented a similar emission spectrum for PSI as those acclimated to the SOL light condition for 3 days and slightly higher values of emission for PSII. Also, PCC6803 samples after 21 days under the different light spectra presented different emission features. SOL samples were identical to those acclimated for 3 days, whereas cultures acclimated for a longer time to M7 presented a higher PSII emission with respect to those tested after 3 days. 21 days FR acclimated samples were instead drastically different from those acclimated for 3 days. The major emission peaks shifted from 725 nm to 685 and 695 nm, and the emission of PSI was reduced roughly by half.

## $O_2$ evolution is very efficient in FaRLiP and non-FaRLiP cyanobacteria acclimated to M7

After the long-term acclimation of the strains, their $O_2$ evolution was assessed for 24 h (Figure 10) directly under the different growth light conditions by using the ASC setup. M7 and SOL acclimated samples of both strains showed a similar $O_2$ evolution efficiency, while differences arose in FR acclimated samples: cells of PCC6912 could evolve $O_2$ in FR, cells of PCC6803 instead were unable to produce any net $O_2$ in FR. These trends confirmed those obtained from the growth curves for each strain and each light condition.

## Discussion

The possibility of Oxygenic Photosynthesis (OP) on terrestrial exoplanets has been discussed for a long time now. OP is the most important bioprocess known on Earth: it shaped the atmosphere we breathe and allowed complex forms of life to evolve and spread on the planet (Gale and Wandel, 2016), and nowadays it produces nearly 100% of Earth's biomass (Covone et al., 2021). Of course, the only OP we are aware of is that arose on Earth, therefore, we don't know how a similar process could have evolved on other planets of the Solar System (if any) or on planets orbiting different kinds of stars. As M-dwarfs are the most abundant stars present in our galaxy and recent surveys showed they can host terrestrial exoplanets in their Habitable Zone (HZ) (Hsu et al., 2020), the question shifted to whether or not OP could be performed in exoplanets orbiting those stars. This is not trivial, since water splitting, a crucial step in OP, requires photons energetic enough to trigger the reaction. These photons in the visible range are abundant for terrestrial-like planets orbiting Sun-like stars but are limited in the same conditions for an M-dwarf spectrum. Photons in the IR instead, abundant in the M-dwarf spectrum, could not be energetic enough to trigger the process if terrestrial OP took place (Kiang et al., 2007). Some researchers believe however OP could work similarly as on Earth (Ritchie et al., 2018; Wandel and Gale, 2020; Covone et al., 2021). It was calculated indeed that on tidally locked exoplanets orbiting the HZ of M-dwarfs, there would be enough photon flux in the visible to support Earth-like photosynthesis and that productivity could range from 13 to 22% to that on Earth (Ritchie et al., 2018; Wandel and Gale, 2020). Also in Kiang et al. (2007) a similar conclusion was made, even if they also speculated on the possibility of performing OP utilizing multiple chained photosystems (e.g. 3 or 4) and photons of longer wavelengths (up to 1400 nm) or more photons per molecule of $O_2$, both popular hypotheses in the scientific community (Wolstencroft and Raven, 2002; Tinetti et al., 2006; Chen and Blankenship, 2011; Takizawa et al., 2017; Lehmer et al., 2018). Lingam et al. (2021) went even further, proposing OP performed through far-red absorbing pigments not belonging to the family of chlorophylls. Wandel and Gale (2020) however disagree on the evolution of far-red absorbing pigments, as they conclude there would be no evolutionary pressure for them to appear since there could be a surplus of light in the visible for M-dwarf planets. As an example, the exoplanet Proxima Centauri b (Anglada-Escudé et al., 2016) orbiting the M-dwarf Proxima Centauri, the closest star to our Solar System (4.2 light-years), experiences an irradiance of about 64 µmol m$^{-2}$ s$^{-1}$ in the visible, approximately the 3% of Earth's irradiance by the Sun. This is more than 20 and 180 times the amount light of required by some land plants and cyanobacteria to survive and grow, respectively (Raven et al., 2000; Nobel, 2009). Moreover, a model based on that stellar environment shows that organisms would be light-limited, but still have gross productivities higher than those obtained for most terrestrial grasslands and the open ocean on Earth (Ritchie et al., 2018). Despite the debate, OP hypotheses for M-dwarf exoplanets are indeed mainly based on theoretical models or argued from studies focusing on ecological niches enriched in far-red light: under forest canopies, in water columns, caves, microbial mats, stromatolites, endolithic environments, or under the sands (Gilbert et al., 2001; Chen et al., 2012; Behrendt et al., 2015; Gan and Bryant, 2015; Wolf and Blankenship, 2019; Zhang et al., 2019; Behrendt et al., 2020; Kühl et al., 2020). This field remains largely unexplored from an experimental point of view, with only one experiment to date (Claudi et al., 2021) carried out by acclimating different cyanobacteria strains growing in solid media to a simulated spectrum of an M-dwarf and assessing both their growth and their photosynthetic efficiency using a PAM fluorometer, thus with indirect measurements of chlorophyll fluorescence parameters. Here we physiologically characterized liquid cultures acclimated to an M7 simulated light spectrum, comparing the responses to FR and SOL light sources, measuring their growth through OD and dry weight, and assessing their photosynthetic apparatus features by recording *in vivo* absorption and 77K fluorescence spectra. Moreover, we measured the $O_2$ evolution of acclimated cultures when directly growing under the simulated M-dwarf, solar or far-red light spectra, thanks to the newly developed setup including the ASC. This was particularly important to evaluate the contribution of complex light spectra (as the simulated M-dwarf one) to oxygenic photosynthesis. The acclimation experiments have been carried out at low light intensity to further reduce the amount of photons in the visible portion of an M-dwarf spectrum (roughly one-sixth the light that would be present on Proxima Centauri b), to assess even unfavorable conditions. We could confirm our preliminary results in solid media cultures (Claudi et al., 2021) and further investigate the acclimation features of selected strains from that study. Strains capable and incapable of FaRLiP alike could acclimate and grow similarly under the simulated M7 and SOL lights, despite the different availability of visible light in the two spectra (about 10 and 26 µmol m$^{-2}$ s$^{-1}$, respectively). As for PCC6912, capable of FaRLiP response, it could be expected a slower growth in M7 with respect to SOL: FaRLiP response would indeed redirect energy from growth to the biosynthetic pathways involved in the synthesis of chlorophylls *d* and *f*, and all the alternative subunits of the major components of the PSI, PSII and PBS, necessary for the reorganization of the photosynthetic apparatus in far-red light (Gan et al., 2014a; Gan et al., 2014b). This was not the case, as optical density and dry weight measurements in M7 at 21 days were comparable to those in SOL (Figures 4A, C). As for PCC6803, incapable of FaRLiP response, much lower growth in M7 than in SOL was expected, as the strain could only rely on the modulation of pigments able to absorb red light but poorly capable of absorbing FR light. It is interesting to note that both strains acclimated to M7 by increasing the concentration of pigments per dry weight absorbing the few photons in the visible (Table 1). This is an acclimation process which has been observed previously under red monochromatic light or low light (Mullineaux, 2001; Bernát et al., 2021). Moreover, PCC6912 activated FaRLiP to grow in M7, even if its activation was slower compared to FR, as seen after 3 days from the *in vivo* spectra (Figure S3). Evidence of FaRLiP activation in M7 could be found eventually through the detection of chlorophylls *d* and *f* at the HPLC at 3 days (Figure S4) and the detection of the rearranging photosynthetic apparatus through 77 K fluorescent measurements (Figure 8). The progression of the acclimation process was confirmed by the same measurements after 21 days (Figures 6A, 7A, 9A). In cyanobacteria, under normal light conditions (e.g. white light, WL) photosystems have 3 peaks of fluorescence when excited at 440 nm, of which two come from PSII antenna pigments (WL-PSI, 685, 695 nm) and one comes from PSI antenna pigments and is variable in wavelength (WL-PSII, 710-735 nm), depending on the species (Murakami, 1997). Conversely, under far-red light conditions (far-red light, FRL), strains capable of activating FaRLiP show a variation in the emission spectra that can be detected. PSI shifts its peak emission from 710-735 nm to roughly 740 nm (FRL-PSI), due to the reorganization of the photosynthetic apparatus and the presence of Chl *f* as an antenna pigment in the PSI (Gan et al., 2014a). Under M7, PCC6912 activated FaRLiP extending the absorption of light in the far-red by synthesizing chlorophylls *d* and *f* and the subunits of the PSI necessary to host them. After 3 days, only a small shift in the PSI emission at 440 nm of cells acclimated to M7 compared to FR could be seen, which meant only a small portion of WL-PSI was substituted with FR-PSI. Conversely, after 21 days, the shift in M7 acclimated samples was identical to that in 3 days FR acclimated samples, meaning that in M7 a similar reorganization of the photosynthetic apparatus to that in FR was taking place at a slower pace. The reorganization of the photosynthetic apparatus allowed PCC6912 to absorb light in M7 and evolve $O_2$ efficiently (Figure 10A). Data from 77K spectroscopy highlighted an acclimation response to FR also in PCC6803. This strain, incapable of FaRLiP, responded to FR by deeply changing the ratio of PSII to PSI (Figure 9B), similarly to what had been seen in the past under long-wavelength red light (Murakami, 1997). This reorganization however didn't lead to any net $O_2$ production in FR light (Figure 10B). On the contrary, though, PCC6803 could exploit efficiently the light in M7 by only adjusting its light-harvesting capabilities, evolving afterwards comparable amounts of $O_2$ in M7 and SOL. Recent studies on plants show that far-red light is used efficiently for oxygenic photosynthesis when in combination with shorter wavelength, while instead is poorly utilized when provided alone (Zhen, 2020; Zhen and Bugbee, 2020; Zhen et al., 2022), even in the absence of exotic acclimations like those performed by cyanobacteria. This has led the authors to question the real capabilities of oxygenic photosynthetic organisms to absorb far-red light and to extend the definition of PAR to a new broader definition, extended PAR (or ePAR, 400 – 750 nm) (Zhen et al., 2021). In this frame then we could hypothesize that PCC6803 is able to grow in M7 similarly to SOL but not in FR, because it can utilise far-red light efficiently only when in combination with visible light, as in the case of M7, and not alone, as in the case of FR. The overall picture that we obtained then, seems to be that cyanobacteria could utilize an M-dwarf light to photosynthesize, without synthesizing new forms of far-red absorbing pigments. For those who are capable of FaRLiP instead, the response is active and seems to be slower in M7 with respect to FR. So, even if in theory OP shouldn't be as efficient under an M-dwarf spectrum as it is under a G-type one like the Sun, in practice Earth-like OP may be a possibility on the surface of those planets and could produce $O_2$ biosignatures detectable remotely if other boundaries conditions are met (Rugheimer and Kaltenegger, 2018; Schwieterman et al., 2018). However, it cannot be excluded that organisms evolved under an M-dwarf spectrum could rely on other types of oxygenic or anoxygenic photosynthesis, which can use a variety of organic and inorganic molecules as donors of electrons (George et al., 2020). Those molecules, in the case of anoxygenic photosynthesis for example, could be hardly available as $H_2O$ is on Earth, thus leading to low global productivities of anaerobic organisms (Kiang et al., 2007). Such life would be thus difficult to detect remotely and biosignatures would be not robust enough (e.g., keen to false positives). Finally, it is important to highlight that the activation of FaRLiP in M7 and its detectability *in vivo* is of interest for the search of other biosignatures, in particular the red-edge (Seager et al., 2005). This is a so-called surface biosignature and is a spectral feature that arises from the reflectance properties of photosynthetic pigments, which absorb light in the visible and reflect it in the infrared, and is a feature displayed at roughly 700 nm only by oxygenic photosynthetic organisms (Seager et al., 2005; Tinetti et al., 2006; Kiang et al., 2007; Takizawa et al., 2017; Cavalazzi and Westall, 2019; Lingam and Loeb, 2021). Given that the *in vivo* absorption and reflectance spectra of oxygenic photosynthetic organisms are related (in fact one is almost the complementary to the other), results obtained for PCC6912 (Figure 6A) imply that, if life utilizing a far-red photoacclimation took place in a distant exoplanet orbiting an M-dwarf star, we could possibly see a shift or a distortion in the Red-Edge feature of the planet, making it easier to distinguish a positive detection from false ones. In conclusion, the objective of this study was to assess the growth, acclimation strategies and photosynthetic rate of selected cyanobacterial strains under a simulated M-dwarf light spectrum. Both cyanobacteria tested (*Chlorogloeopsis fritschii* PCC6912, able to activate the FaRLiP response, and *Synechocystis* sp PCC6803, unable to perform such acclimation) could exploit the simulated light conditions to grow and photosynthesize efficiently. *C. fritschii* PCC6912 by activating FaRLiP and expanding its capabilities for oxygenic photosynthesis beyond 700 nm, *Synechocystis* sp. PCC6803 possibly by utilizing far-red light efficiently only in combination with visible light, even if present in a very low amount as in the simulated M7 spectrum. Both acclimations led to efficient $O_2$ evolution, showing the potentiality of cyanobacteria to utilize light regimes that could arise on tidally locked planets orbiting the HZ of M-dwarf stars and to produce potentially detectable $O_2$ biosignatures detectable from remote. Finally, FaRLiP activation in *C. fritschii* PCC6912 under M-dwarf simulated light conditions opens up to the detectability of peculiar surface biosignatures produced by far-red acclimated oxygenic photosynthetic organisms on the surface of these exoplanets, increasing our chances to detect life beyond our Solar System.

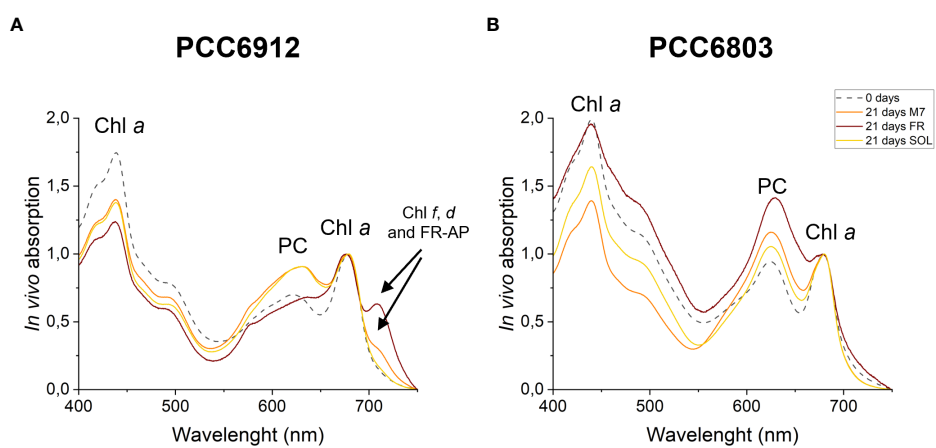

FIGURE 6
In vivo absorption spectra of PCC6912 (A) and PCC6803 (B) after 21 days. Where present, absorption from pigments involved in the FaRLiP response is highlighted with a black arrow. M7, M-dwarf light; FR, Far-red light; SOL, Solar light; Chl a, Chlorophyll a; PC, Phycocyanin; Chl f, Chlorophyll f; Chl d, Chlorophyll d; FR-AP, far-red induced allophycocyanin. Spectra are normalized at 680 nm to highlight FaRLiP.

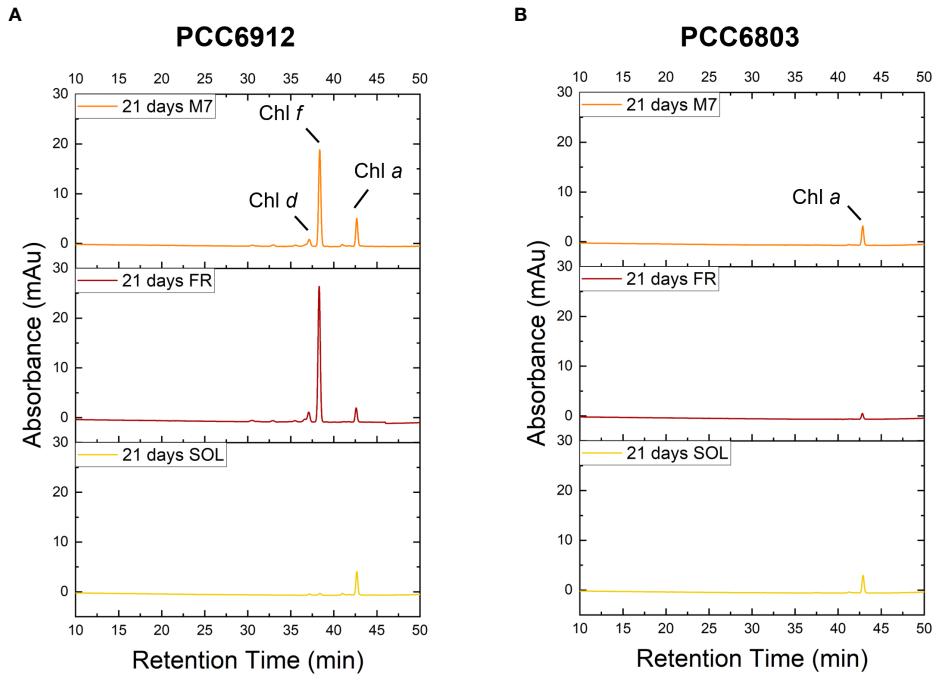

FIGURE 7
HPLC Chromatogram at 705 nm of PCC6912 **(A)** and PCC6803 **(B)** samples after 21 days of exposure. M7, M-dwarf light; FR: Far-red light; SOL, Solar light; Chl *a*, Chlorophyll *a*; Chl *f*, Chlorophyll *f*; Chl *d*, Chlorophyll *d*.

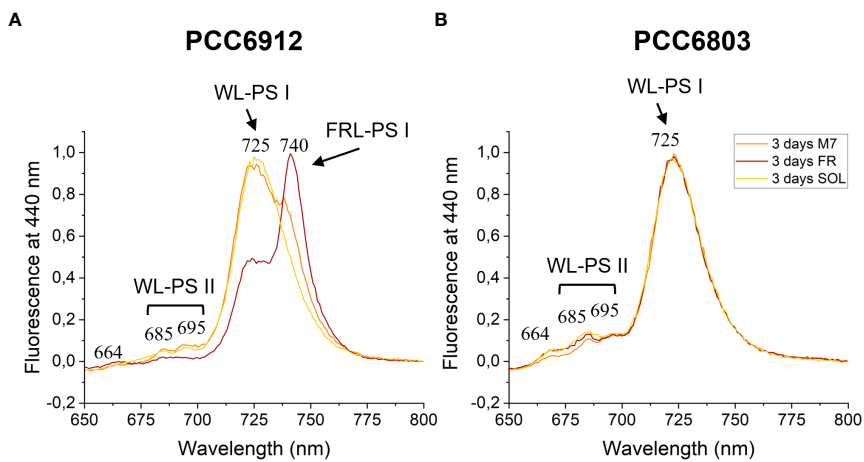

FIGURE 8
77 K fluorescence emission spectra of PCC6912 **(A)** and PCC6803 **(B)** samples after 3 days of exposure. Excitation light was set at 440 nm. Spectra are normalized to the maximum peak of fluorescence of each sample. WL-PSI, WL-PSII and FR-PSI respectively designate emission fluorescence of white light photosystem I, white light photosystem II and far-red light photosystem I; M7, M-dwarf light; FR, Far-red light; SOL, Solar light.

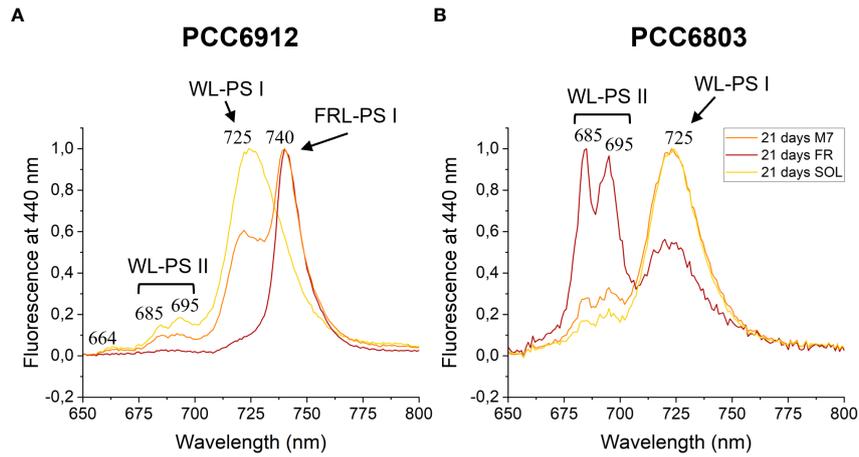

FIGURE 9
77 K fluorescence emission spectra of PCC6912 (A) and PCC6803 (B) samples after 21 days of exposure. Excitation light was set at 440 nm. Spectra are normalized to the maximum peak of fluorescence of each sample. WL-PSI, WL-PSII and FRL-PSI respectively designate emission fluorescence of white light photosystem I, white light photosystem II and far-red light photosystem I; M7, M-dwarf light; FR, Far-red light; SOL, Solar light.

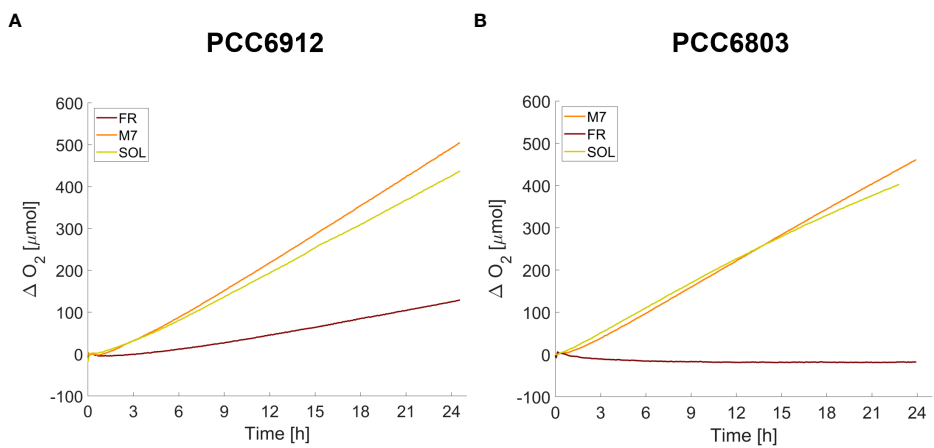

FIGURE 10
$O_2$ traces of PCC6912 (A) and PCC6803 (B) strains acclimated for 21 days to the different light spectra. T$_0$ was set at 3.600 s to exclude the initial gas equilibration period inside the ASC.

## Data availability statement

The raw data supporting the conclusions of this article will be made available by the authors, without undue reservation.

## Author contributions

NL, RC and LP made the conceptualisation and supervision of the whole work. MB, NL, and LC conceived the Solar Simulator and the Far-red Simulator. NL, RC, LP, LC and MB conceived the Atmosphere Simulator Chamber (ASC), LC realized it. LC and MB assembled and tested the ASC setup. NL, MB and TM conceived the biological experiments, MB, AP, AS, DS performed and analysed them, NL, TM and LC helped in the analysis. MB and NL wrote the manuscript, all authors critically read it. All authors contributed to the article and approved the submitted version.

## Funding

The research was co-funded by the Italian Space Agency through the "Life in Space" project (ASI N. 2019-3-U.0), and by the Department of Biology of the University of Padova and the Institute for Photonics and Nanotechnologies of CNR through intramural grants.

## Conflict of interest

The authors declare that the research was conducted in the absence of any commercial or financial relationships that could be construed as a potential conflict of interest.

## Publisher's note

All claims expressed in this article are solely those of the authors and do not necessarily represent those of their affiliated organizations, or those of the publisher, the editors and the reviewers. Any product that may be evaluated in this article, or claim that may be made by its manufacturer, is not guaranteed or endorsed by the publisher.

## Supplementary material

The Supplementary Material for this article can be found online at: https://www.frontiersin.org/articles/10.3389/fpls.2023.1070359/full#supplementary-material